# Group Authentication Using The Naccache-Stern Public-Key Cryptosystem

*Scott Guthery*
*sguthery@mobile-mind.com*

*Abstract*

A group authentication protocol authenticates pre-defined groups of individuals such that:

- No individual is identified
- No knowledge of which groups can be successfully authenticated is known to the verifier
- No sensitive data is exposed

The paper presents a group authentication protocol based on splitting the private keys of the Naccache-Stern public-key cryptosystem in such a way that the Boolean expression defining the authenticable groups is implicit in the split.

## Introduction

Acme Inc. has a corporate airplane. The Acme corporate plane can only be used by Acme employees. Any trip must include at least two employees and at least one must be a manager. On both privacy and security grounds Acme does not want to identify which individuals are aboard any trip on the airplane.

Suppose Acme has five employees. A and B are managers while C, D, and E are non-managers. Acme's airplane usage policy can be expressed as follows:

(A and B) or ((A or B) and (C or D or E))

To implement its usage and privacy policy Acme splits a private key among the employees in a way that only groups of employees that satisfy this Boolean expression can use the plane. Acme puts each key share on a separate security token.

When a group of individuals arrives at the airport, they present their smart cards to an untrusted verifier. The verifier generates a random challenge, encrypts it with Acme's public key and provides the result to each of the tokens. Each token process the data it is given and returns a response to the verifier. The verifier combines the responses from the tokens and if the combination equals the original random number, the group of individual is permitted to use the Acme airplane. Otherwise, the group is not permitted to use the Acme airplane.

## Group Authentication

This example illustrates three desirable properties of a group authentication protocol:

1. A group can be authenticated as a unit without authenticating any individual member of the group.
2. The verifier need not know the definition of the group or groups that can be successfully authenticated.
3. The authentication can be performed by an untrusted entity without reconstructing any sensitive data.

The particular application circumstance that motivated this work is a multi-entity transaction system wherein we wish to control which entities can co-operate in forming a transaction without identifying individual entities and without handling any sensitive data.

*Previous Work*

There are many key splitting and key sharing schemes. An extensive bibliography is provided by Stinson and Ruizhong [9] and an overview by Seberry et al [8]. Benaloh and Leichter [2] show that keys can be split and shared so that only groups satisfying an arbitrary AND/OR Boolean expression can reconstruct the secret key. There are threshold (k-of-n) key sharing schemes that enable group authentication without recreating the secret key, for example the protocol of Suida, Freund and Huang [6]. Representing key shares as subsequences of binary sequences has been considered by number of authors including Anderson [1], Davida [3], Massey[4], and McEliece and Sarwate [5].

*The Naccache-Stern Public-Key Cryptosystem*

Naccache and Stern [6] present a public-key cryptosystem based on the knapsack problem. We present herein a method whereby the private key of the Naccache-Stern system can be shared among a set of key holders in a way that satisfies the above properties of a group authentication protocol.

The decryption function of the Naccache-Stern cryptosystem is given by:

$$m = \sum \{ \ 2^i \ | \ p_i \text{ is a factor of } c^s \bmod p \ \}$$

where $P = \{p_i\}$ is a set of prime numbers, p is a prime greater than $\Pi \ p_i$, c is the ciphertext and s is the secret key.

*Naccache-Stern Key Splitting Relative to an AND/OR Boolean Expression*

Let $A = \{A_j\}$ be a set of key holders and let B be an AND/OR Boolean expression over A. For example, if $A = \{A_1, A_2, A_3\}$ then B might be

$$(A_1 \text{ and } A_2) \text{ or } (A_1 \text{ and } A_3)$$

Apply the algorithm of Benaloh and Leichter [2] to partition the set P into parts $P_j$ according to B. In the language of [1], a set is moved across an AND operator in B by partitioning the set arbitrarily into non-null subsets and across an OR operator in B by duplicating the set. At the conclusion of the algorithm, set $P_j$ to the union of all sets associated with the variable $A_j$. $P_j$ is non-null for all j since without loss of generality each $A_j$ is mentioned at least once in B.

The tuple $(P_j, s)$ is $A_j$'s share of the private key. $A_j$'s contribution, $m_j$, to the decryption of a ciphertext c is given by

$$m_j = \sum \{ 2^i \mid p_i \in P_j \text{ and } p_i \text{ is a factor of } c^s \bmod p \}$$

and the plaintext message m is given by

$$m = \sum m_j$$

where we take $\sum$ to be bitwise non-exclusive logical OR.

*A Small Example*

Suppose A = $\{A_1, A_2, A_3\}$ are three entities that according to the governing policy may only co-operate according to the Boolean expression

$$(A_1 \text{ and } A_2) \text{ or } (A_1 \text{ and } A_3)$$

That is, $A_1$ and $A_2$ may co-operate and $A_1$ and $A_3$ may co-operate but $A_2$ and $A_3$ may not co-operate in the absence of $A_1$ nor can any of them act alone.

When any group of entities attempt to co-operate, the untrusted verifier generates a random message m, encrypts it using the public key of the Naccache-Stern cryptosystem, and sends the resulting ciphertext c to each entity as a challenge. If and only if the responses from the entities are such that

$$m = \sum m_j$$

are the entities are allowed to co-operate.

For a specific example, consider the public key cryptosystem given as "a small example" in the Naccache-Stern paper [6]. For this system, n = 7, P = $\{2, 3, 5, 7, 11, 13, 17, 19\}$, and s = 5,642,069. If the random message m is 202, then the challenge c computed using the public key and provided to each token is

$$c = \prod v_i^{x_i} \bmod p = 7202882$$

where

$$m = \sum 2^{x_i}$$

Suppose the secret key has been divided among the three entities as follows:

$$P_1 = \{2, 3, 5, 7\} \text{ and } P_2 = P_3 = \{11, 13, 17, 19\}$$

Then

$$m_1 = \sum \{ 2^i \,|\, p_i \in \{ 2, 3, 5, 7\} \text{ and } p_i \text{ is a factor of } 202\} = 2^3 \vee 2^1$$

and

$$m_2 = m_3 = \sum \{ 2^i \,|\, p_i \in \{11, 13, 17, 19\} \text{ and } p_i \text{ is a factor of } 202\} = 2^7 \vee 2^6$$

*Non-Monotonic Access Structures*

Returning to the example in the first section, it should be noted that application of the Benaloh and Leichter algorithm to Acme's Boolean expression does not enforce the restriction imposed by the seating capacity of the airplane. The group {A, B, C, D} for example would be successfully authenticated using above formulation.

The Benaloh and Leichter algorithm as well as much of the work on shared keys considers only monotone access structures; viz. if a subset of A' of A is allowed and A' $\subset$ A'', then A'' is allowed. In some group authentication contexts this monotonicity condition is not acceptable. We may, for example, wish to allow an entity to form a transaction in co-operation with some entities but not in the concurrent presence of others.

One method of realizing non-monotone access structures; that is structures defined by Boolean expressions including the NOT operator as well as AND and OR; provides each key holder with an ordered sequence of key shares and retrieves an ordered sequence of responses to the challenge. The $i^{th}$ response of a key holder j to the challenge, $r_{ij}$ is computed using the key holder j's $i^{th}$ share.

For non-monotonic access structures, we take the response merging function $\sum$ to be the arithmetic sum or XOR of the key holder responses rather than the non-exclusive OR. The overall response r is given by

$$r = (r_1, r_2, \ldots, r_n) = (\sum r_{1j}, \sum r_{2j}, \ldots, \sum r_{nj})$$

If $r_i = m$ for any i, then the group is successfully authenticated.

*Another Example*

In the corporate airplane example, the groups we wish to successfully authenticate are AB, AC, AD, AE, ACD, ABC, ABD, ABE, ACE, ADE, BC, BD, BE, BCD, BCE and BDE.

To form a Naccache-Stern public-key cryptosystem we take

$$P = \{2, 3, 5, 7, 11, 13, 17, 19, 23, 29, 31, 37\},$$

$$p = 7420738134871,$$

and set the private key, s, to

$$s = 5642069$$

The public key of this cryptosystem is given by

$$v[0] = 1042080239371 \qquad v[6] = 6408801185994$$
$$v[1] = 6961378167419 \qquad v[7] = 6664307396372$$
$$v[2] = 556387338943 \qquad v[8] = 6792283659586$$
$$v[3] = 6467374518496 \qquad v[9] = 4009453191992$$
$$v[4] = 6101909563954 \qquad v[10] = 4858036635332$$
$$v[5] = 7161849266528 \qquad v[11] = 3535089085276$$

Table 1 shows the key share sequence placed on each employee's token to authenticate the 13 subsets of {A, B, C, D, E} needed for the corporate airplane example.

| Sequence Index | A | B | C | D | E | Group(s) Authenticated |
|---|---|---|---|---|---|---|
| 1 | 2, 3, 5, 7, 11, 13 | 2, 3, 5, 7, 11, 13 | 17, 19, 23, 29, 31, 37 | 17, 19, 23, 29, 31, 37 | 17, 19, 23, 29, 31, 37 | AC, AD, AE, BC, BD, BE |
| 2 | 2, 3, 5, 7 | 11, 13, 17, 19 | 23, 29, 31, 37 | 23, 29, 31, 37 | 23, 29, 31, 37 | ABC, ABD, ABE |
| 3 | 2, 3, 5, 7 | | 11, 13, 17, 19 | 23, 29, 31, 37 | 23, 29, 31, 37 | ACD, ACE |
| 4 | | 2, 3, 5, 7 | 11, 13, 17, 19 | 23, 29, 31, 37 | 23, 29, 31, 37 | BCD, BCE |
| 5 | 2, 3, 5, 7 | | | 11, 13, 17, 19 | 23, 29, 31, 37 | ADE |
| 6 | | 2, 3, 5, 7 | | 11, 13, 17, 19 | 23, 29, 31, 37 | BDE |
| 7 | 2, 3, 5, 7, 11, 13 | 17, 19, 23, 29, 31, 37 | | | | AB |

Table 1. Key Share Sequences Held by Each Employee

The encryption of the random number 2919 = 101101100111 using the public key is m = 1073741824.  The series of responses from each employee smart card is given in Table 2.

| Sequence Index | A | B | C | D | E | Groups(s) Authenticated |
|---|---|---|---|---|---|---|
| 1 | 39 | 39 | 2880 | 2880 | 2880 | AC, AD, AE, BC, BD, BE |
| 2 | 7 | 96 | 2816 | 2816 | 2816 | ABC, ABD, ABE |
| 3 | 7 | 1 | 96 | 2816 | 2816 | ACD, ACE |
| 4 | 1 | 7 | 96 | 2816 | 2816 | BCD, BCE |
| 5 | 7 | 1 | 1 | 96 | 2816 | ADE |
| 6 | 1 | 7 | 1 | 96 | 2816 | BDE |
| 7 | 39 | 1 | 2880 | 1 | 1 | AB |

Table 2. Response Sequences from Each Employee Token

Variations of this basic approach include using a different random number for each response in the sequence and having the null response of a key holder be a non-zero random number rather than simply 1.

*Summary*

A method for sharing the private key of a Naccache-Stern public-key cryptosystem among a set of key holders such that only subsets satisfying an arbitrary Boolean expression can decrypt a message encrypted with the system's public key has been described.  The decryption of a message encrypted with the public key of the system can be obtained without identifying any member of the group, without revealing the Boolean expression defining the group, and without bringing the private key into being.

The method is useful for enforcing security policies expressed in terms Boolean expressions over groups of entities.  It provides implicit and simultaneous authentication of the group members together with the evaluation of the policy statement governing their co-operation.